\documentclass[eqsecnum,showpacs,showkeys,nofootinbib,aps]{revtex4}
\usepackage{xcolor}
\usepackage{ulem}\renewcommand{\theequation}{\arabic{equation}}

\begin{document}
\title{Phenomenologies in hypersphere soliton and stringy photon models}
\author{Soon-Tae Hong}
\email{galaxy.mass@gmail.com}
\affiliation{Center for Quantum Spacetime and Department of Physics, Sogang University, Seoul 04107, Korea}
%\date{August 21, 2023}
\date{\today}

\begin{abstract}
We consider the Dirac quantization in the first class formalism to investigate the hypersphere soliton model (HSM) defined on the 
$S^{3}$ hypersphere. To do this, we construct the first class Hamiltonian possessing the Weyl ordering correction. In the HSM, 
we evaluate the baryon physical quantities such as the baryon masses, magnetic moments, 
axial coupling constant and charge radii, most predicted values of which are in good agreement with the corresponding experimental data. 
Moreover, shuffling the baryon and transition magnetic moments, we find the model independent sum rules. 
In the HSM we also evaluate the baryon intrinsic frequencies such as $\omega_{N}=0.87\times 10^{23}~{\rm sec}^{-1}$ and 
$\omega_{\Delta}=1.74\times 10^{23}~{\rm sec}^{-1}$ of the nucleon and delta baryon, respectively, to yield the identity 
$\omega_{\Delta}=2\omega_{N}$. Next, making use of the Nambu-Goto string action and its extended rotating bosonic string theory, 
we formulate the stringy photon model to obtain the energy of the string configuration, which consists of the rotational and vibrational 
energies of the open string. Exploiting this total string energy we evaluate the photon intrinsic frequency 
$\omega_{\gamma}=9.00\times 10^{23}~{\rm sec}^{-1}$ which is comparable to the corresponding baryon intrinsic frequencies.
We also predict the photon size $\langle r^{2}\rangle^{1/2}(\rm photon)=0.17~{\rm fm}$ which is approximately 21\% of 
the proton magnetic charge radius. 
\end{abstract}


\maketitle

%%%%%%%%%%%%%%%%%%%%%%%%%%%%%%%%%%%%%%%%%%%%%%%%%%%%%%%%%%%%%%%%%%%%%%%%
\section{Introduction}
\renewcommand{\theequation}{\arabic{section}.\arabic{equation}}
%%%%%%%%%%%%%%%%%%%%%%%%%%%%%%%%%%%%%%%%%%%%%%%%%%%%%%%%%%%%%%%%%%%%%%%%

In this work we will consider the extended objects instead of the point particles. It is well known that, 
as the extended objects, we have the solitons~\cite{skyrme61,faddeev,anw83,manton1,liu87,hong98plb,manton2,hong15,hong21} 
and strings~\cite{green87,gova,polchinski98,hong22}. 
In particular, in this review we will mainly use the hypersphere soliton model (HSM)~\cite{manton1,hong98plb,hong21} and stringy photon 
model (SPM)~\cite{hong22}. To be more specific, in the soliton models, 
we have the standard Skyrmion which describes baryon static properties in $R^{3}$ manifold~\cite{skyrme61,anw83,liu87,hong15}. This 
model was proposed by Skyrme in 1961~\cite{skyrme61}. 
In this paper we will consider the paper by Adkins, Nappi and Witten (ANW)~\cite{anw83,liu87}, to compare with the HSM. 
Next we will investigate the HSM which is formulated on the hypersphere $S^{3}$ instead of $R^{3}$~\cite{manton1,hong98plb,hong21}. 
Exploiting the HSM, we will evaluate the baryon physical quantities most of which are in good agreement with 
the corresponding experimental data.

In 1962 the electron was proposed as a charged conducting surface by Dirac~\cite{di62}. 
According to his proposal, the electron shape and size should pulsate. Here the surface tension of the 
electron was supposed to prevent the electron 
from flying apart under the repulsive forces of the charge. Motivated by his idea, 
we will investigate pulsating baryons in the first class formalism in the HSM~\cite{manton1,hong98plb}, 
to evaluate the intrinsic frequencies of the baryons, baryon masses with the Weyl ordering correction (WOC)
and axial coupling constant~\cite{hong21}. 

On the other hand, as regards the string theories, we have the critical higher dimensional string 
theory~\cite{green87,gova,polchinski98}, and the recently proposed SPM defined in the four dimensional spacetime
which predicts the photon radius, and the photon intrinsic frequency comparable to the corresponding baryon 
intrinsic frequencies~\cite{hong22}. In the SPM we have exploited the open string which 
performs both rotational and vibrational motions~\cite{hong22}. Note that the rotational degrees of freedom of the 
photon have been investigated in the early universe~\cite{hongcho08,hongcho11}. 

In this review, we will exploit the HSM in the first class Dirac Hamiltonian formalism, 
to evaluate the physical quantities such as the baryon masses, magnetic moments, axial 
coupling constant, charge radii and baryon intrinsic frequencies. Next in the SPM we will predict the 
photon intrinsic frequency which is shown to be comparable to the baryon intrinsic ones. To do this, we will exploit 
the Nambu-Goto string theory~\cite{nambu70,goto71}. In the SPM we will next introduce an open string action associated 
with the photon~\cite{schwarz74}. 
%--------------to add
Making use of the rotational and vibrational energies of the string, 
we will evaluate explicitly the photon intrinsic frequency with which, assuming that the photon size is given by 
the string radius in the SPM, we will predict the photon size.
%--------------

In Sec. 2, we will predict the baryon properties in the HSM. In Sec. 3, we will evaluate the intrinsic frequencies of the 
baryons in the HSM. In Sec. 4, we will exploit the SPM to predict the photon intrinsic frequency and photon size. 
Sec. 5 includes conclusions.

%%%%%%%%%%%%%%%%%%%%%%%%%%%%%%%%%%%%%%%%%%%%%%%%%%%%%%%%%%%%%%%%%%%%%%%%
\section{Baryon Predictions in HSM}
\setcounter{equation}{0}
%\theequation
\renewcommand{\theequation}{\arabic{equation}}
%\renewcommand{\theequation}{\arabic{section}.\arabic{equation}}
%%%%%%%%%%%%%%%%%%%%%%%%%%%%%%%%%%%%%%%%%%%%%%%%%%%%%%%%%%%%%%%%%%%%%%%%

Now we consider the baryon predictions in the first class Hamiltonian formalism in the HSM. 
To do this, we introduce the Skyrmion Lagrangian density given by  
\begin{equation}
{\cal L}=\frac{f_{\pi}^{2}}{4}{\rm tr}(\partial_{\mu}U^{\dagger}
         \partial^{\mu}U)+\frac{1}{32e^{2}}{\rm tr}[U^{\dagger}\partial_{\mu}U,
         U^{\dagger}\partial_{\nu}U]^{2}
\label{sklag}
\end{equation}
where $U$ is an SU(2) chiral field, and  $f_{\pi}$ and $e$ are a pion decay constant
and a dimensionless Skyrme parameter, respectively. In this work, we will treat $f_{\pi}$ and $e$ as the model parameters. 
Here the quartic term is necessary to stabilize the soliton in the baryon sector.  

Next we introduce the hyperspherical three metric on $S^{3}$ of the form
\begin{equation}
ds^{2}=\lambda^{2}d\mu^{2}+\lambda^{2}\sin^{2}\mu~(d\theta^{2}+\sin^{2}\theta~d\phi^{2}),
\label{ds2}
\end{equation}
where the ranges of the three angles are defined as $0\le\mu\le\pi$, $0\le\theta\le\pi$ and $0\le\phi\le 2\pi$, 
and $\lambda$ ($0\le \lambda<\infty$) is a radius parameter of $S^{3}$. In the HSM, using the Skyrmion Lagrangian density in 
(\ref{sklag}) we obtain the soliton energy $E$ of the form
\begin{equation}
E=\frac{f_{\pi}}{e}\left[2\pi L\int_{0}^{\pi}d\mu\sin^{2}\mu\left(\left(\frac{d f}{d\mu}
    +\frac{1}{L}\frac{\sin^{2}f}{\sin^{2}\mu}\right)^{2}+2\left(\frac{1}{L}\frac{d f} 
     {d\mu}+1\right)^{2}\frac{\sin^{2}f}{\sin^{2}\mu}\right)+6\pi^{2}B\right],
\label{solenergy}
\end{equation}
where $L=ef_{\pi}\lambda$ ($0\le L<\infty$) is a dimensionless radius parameter and $B$ is topological 
baryon number, which is unity for a single soliton. Here $f(\mu)$ is a profile function for hypersphere soliton, and 
satisfies $f(0)=\pi$ and $f(\pi)=0$ for unit topological baryon number. 
Note that the the profile function $f$ in the soliton energy $E$ in (\ref{solenergy}) satisfies the first order differential equations
\begin{equation}
\frac{d f}{d\mu}+\frac{1}{L}\frac{\sin^{2}f}{\sin^{2}\mu}=0,~~~\frac{1}{L}\frac{d f}{d\mu}+1=0,
\label{diff}
\end{equation}  
to attain the BPS topological lower bound in the soliton energy~\cite{manton1,manton2,hong98plb,faddeev,liu87} given by 
\begin{equation}
E_{B}=\frac{6\pi^{2}f_{\pi}}{e}.
\label{bpsbound}
\end{equation}
Moreover, in this case we find the equation of motion for the hypersphere soliton~\cite{manton1,hong98plb}
\begin{equation}
\left(L\sin^{2}\mu+\frac{2}{L}\sin^{2}f \right)\frac{d^{2}f}{d\mu^{2}}
 +\left(L\sin 2\mu+\frac{1}{L}\frac{d f}{d\mu}\sin 2f\right)\frac{d f}{d\mu}
 -\left(L+\frac{1}{L}\frac{\sin^{2}f}{\sin^{2}\mu}\right)\sin 2f=0.
\label{eom22}
\end{equation}
One of the simplest solution of (\ref{eom22}) is the identity map
\begin{equation}
f(\mu)=\pi-\mu,
\label{identitymap1}
\end{equation}
in which case the soliton energy in (\ref{solenergy}) can be rewritten as~\cite{manton1,hong98plb} 
\begin{equation}
E=\frac{3\pi^{2}f_{\pi}}{e}\left(L+\frac{1}{L}\right).
\label{solitonenergy22}
\end{equation}
Note that, in order to obtain the BPS topological {\it lower bound} $E_{B}$ in (\ref{bpsbound}) by exploiting the soliton energy $E$ 
in (\ref{solitonenergy22}) associated with the identity map $f(\mu)=\pi-\mu$ in (\ref{identitymap1}), we use 
the fixed value $L=L_{B}$ where 
\begin{equation}
L_{B}\equiv ef_{\pi}\lambda_{B}=1.\label{lbequiv} 
\end{equation}
%--------------to add
Note also that the identity map in (\ref{identitymap1}) is a minimum energy solution for $L<\sqrt{2}$, 
while for $L>\sqrt{2}$ it is a saddle point~\cite{manton3,jackson88}.
%--------------

Now we briefly discuss the Dirac quantization of constrained system~\cite{skyrme61,anw83,manton1,manton2,liu87,hong98plb,hong15,hong21,di}. 
In the HSM, we have the second class constraints 
for the collective coordinates $a^{\mu}$ ($\mu=0,1,2,3$) 
and the corresponding canonical momenta $\pi^{\mu}$ conjugate to $a^{\mu}$ of the form
\begin{equation}\Omega_{1}=a^{\mu}a^{\mu}-1\approx 0,~~~\Omega_{2}=a^{\mu}\pi^{\mu}\approx 0.
\label{secondconst}
\end{equation}
\textcolor{black}{Exploiting the Poisson bracket for $a^{\mu}$ and $\pi^{\mu}$, 
\begin{equation}\{a^{\mu},\pi^{\nu}\}=\delta^{\mu\nu},
\label{amupimu}
\end{equation}}
we obtain the Poisson algebra for the commutator of $\Omega_{1}$ and $\Omega_{2}$~\cite{di,hong15}
\begin{equation}
\{\Omega_{1},\Omega_{2}\}=2a^{\mu}a^{\mu}\neq 0.
\label{o1o2poi}
\end{equation}
Since this Poisson algebra does not vanish we call the constraints $\Omega_{1}$ and $\Omega_{2}$, the second class.

In the HSM, spin and isospin states can be treated
by collective coordinates $a^{\mu}=(a^{0},\vec{a})$ $(\mu=0,1,2,3)$, corresponding to the spin and isospin 
collective rotation $A(t)\in$ SU(2) given by $A(t) = a^{0}+i\vec{a}\cdot\vec{\tau}$. Exploiting the coordinates $a^{\mu}$, 
we obtain the Hamiltonian of the form
\begin{equation}
H=E_{B}+\frac{1}{8{\cal I}}\pi^{\mu}\pi^{\mu},
\label{hamil61}
\end{equation}
where $\pi^{\mu}$ are canonical momenta conjugate to the collective coordinates $a^{\mu}$. Here the soliton energy lower bound $E_{B}$ is 
given by (\ref{bpsbound}) and moment of inertia ${\cal I}$ is given by 
\begin{equation}
{\cal I}=\frac{3\pi^{2}}{e^{3}f_{\pi}}.
\label{calipion}
\end{equation}
Note that the identity map $f(\mu)=\pi-\mu$ in (\ref{identitymap1}) with condition $L=L_{B}$, where $L_{B}$ is given by (\ref{lbequiv}), 
is used to predict 
the physical quantities such as the moment of inertia ${\cal I}$ in  (\ref{calipion}), baryon masses, 
charge radii, magnetic moments, axial coupling constant $g_{A}$ and intrinsic pulsation frequencies $\omega_{I}$ in the HSM. 
Note also that the hypersphere coordinates $(\mu,\theta,\phi)$ are integrated out in (\ref{solenergy}), 
and $E_{B}$ in (\ref{bpsbound}) is a function of $\lambda_{B}=\frac{1}{ef_{\pi}}$ or equivalently $f_{\pi}$ and $e$ only. 
Similarly, after integrating out the hypersphere coordinates $(\mu,\theta,\phi)$, the physical quantities in (\ref{calipion}), 
(\ref{mn}) and (\ref{mnmdelta})--(\ref{chargeradiiall}) and (\ref{omegai}) are formulated in terms of  
$f_{\pi}$ and $e$ only.

After performing the canonical quantization in the second class formalism in the HSM, we now construct the Hamiltonian spectrum
\begin{equation}
\langle H\rangle=E_{B}+\frac{1}{2{\cal I}}I(I+1),
\label{canh0}
\end{equation}
where $I$ $(=1/2,~3/2,...)$ are baryon isospin quantum numbers. Exploiting  (\ref{canh0}) 
we find the nucleon mass $M_{N}$ for $I=1/2$ and delta baryon mass $M_{\Delta}$ for $I=3/2$, respectively~\cite{hong98plb,hong21}
\begin{equation}
M_{N}=ef_{\pi}\left(\frac{6\pi^{2}}{e^{2}}+\frac{e^{2}}{8\pi^{2}}\right),~~~
M_{\Delta}=ef_{\pi}\left(\frac{6\pi^{2}}{e^{2}}+\frac{5e^{2}}{8\pi^{2}}\right).
\label{mn}
\end{equation}

Next we formulate the first class constraints $\tilde{\Omega}_{1}$ and $\tilde{\Omega}_{2}$ by adding the terms related with 
the St\"uckelberg fields $\theta$ and $\pi_{\theta}$
\begin{equation}\tilde{\Omega}_{1}=a^{\mu}a^{\mu}-1+2\theta=0,~~~
\tilde{\Omega}_{2}=a^{\mu}\pi^{\mu}-a^{\mu}a^{\mu}\pi_{\theta}=0.
\label{tilde12}
\end{equation}
Here $\theta$ and $\pi_{\theta}$ satisfy the following Poisson bracket
\begin{equation}\{\theta,\pi_{\theta}\}=1,
\label{augamupimu}
\end{equation}
to produce the first class Poisson algebra for the first class constraints $\tilde{\Omega}_{1}$ and $\tilde{\Omega}_{2}$
\begin{equation}\{\tilde{\Omega}_{1},\tilde{\Omega}_{2}\}=0.
\label{tildeo1o2}
\end{equation}
%--------------to add
Now we investigate the operator ordering problem in the first class Hamiltonian formalism. To do this, 
we construct the first class Hamiltonian~\cite{hong21}
\begin{equation}
\tilde{H}=E_{B}+\frac{1}{8{\cal I}}(\pi^{\mu}-a^{\mu}\pi_{\theta})(\pi^{\mu}
-a^{\mu}\pi_{\theta})\frac{a^{\nu}a^{\nu}}{a^{\nu}a^{\nu}+2\theta}.
\label{firsth0}
\end{equation}
Applying the first class constrains in (\ref{tilde12}) to (\ref{firsth0}), we find
\begin{equation}
\tilde{H}=E_{B}+\frac{1}{8{\cal I}}(a^{\mu}a^{\mu}\pi^{\nu}\pi^{\nu}-\pi^{\mu}a^{\mu}a^{\nu}\pi^{\nu}).
\label{firsth2}
\end{equation}
Next we introduce the Weyl ordering procedure~\cite{lee81} to obtain the Weyl ordered operators
\begin{eqnarray}
(a^{\mu}a^{\mu}\pi^{\nu}\pi^{\nu})_{W}^{op}&=&\frac{1}{4}[a^{\mu}(a^{\mu}\pi^{\nu}+\pi^{\nu}a^{\mu})\pi^{\nu}
+\pi^{\nu}(a^{\mu}\pi^{\nu}+\pi^{\nu}a^{\mu})a^{\mu}]
=-\frac{1}{4}(4a^{\mu}a^{\mu}\partial_{\nu}^{2}+8a^{\mu}\partial_{\mu}+3\delta_{\mu\mu}),\nonumber\\
(\pi^{\mu}a^{\mu}a^{\nu}\pi^{\nu})_{W}^{op}&=&\frac{1}{4}(\pi^{\mu}a^{\mu}+a^{\mu}\pi^{\mu})(a^{\nu}\pi^{\nu}
+\pi^{\nu}a^{\nu})=-\frac{1}{4}(4a^{\mu}a^{\nu}\partial_{\mu}\partial_{\nu}+20a^{\mu}\partial_{\mu}+\delta_{\mu\mu}\delta_{\nu\nu}),
\label{weylsym}
\end{eqnarray}
where we have used the quantum operator $\pi^{\mu}=-i\frac{\partial}{\partial a^{\mu}}\equiv -i\partial_{\mu}$. 
Inserting (\ref{weylsym}) into (\ref{firsth2}), we arrive at the Weyl ordered first class Hamiltonian operator
\begin{equation}
\tilde{H}_{W}^{op}=H^{op}+\frac{1}{32{\cal I}}\delta_{\mu\mu}(\delta_{\nu\nu}-3),
\label{firsth3}
\end{equation}
where $H^{op}$ is the second class Hamiltonian operator given by
\begin{equation}
H^{op}=E_{B}+\frac{1}{8{\cal I}}(-a^{\mu}a^{\mu}\partial_{\nu}^{2}+3a^{\mu}\partial_{\mu}+a^{\mu}a^{\nu}\partial_{\mu}\partial_{\nu}).
\label{secondhop}
\end{equation}
Here the last three terms are the three-sphere Laplacian given in terms of the collective 
coordinates and their derivatives to yield the eigenvalues $4I(I+1)$~\cite{neto95}. 
Inserting the relation $\langle H^{op}\rangle=E_{B}+\frac{1}{2{\cal I}}I(I+1)$, 
which is also given in (\ref{canh0}), and the identity $\delta_{\mu\mu}=4$ into (\ref{firsth3}) we construct the Hamiltonian spectrum with 
the WOC in the first class formalism~\cite{hong21}
%--------------
\begin{equation}
\langle \tilde{H}\rangle=E_{B}+\frac{1}{2{\cal I}}\left[I(I+1)+\frac{1}{4}\right],
\label{1stham}
\end{equation}
where $E_{B}$ is the soliton energy BPS lower bound in (\ref{bpsbound}) and ${\cal I}$ is the moment of inertia in 
(\ref{calipion}). 
Comparing the canonical quantization spectrum result $\langle H\rangle$ in (\ref{canh0}) with 
$\langle \tilde{H}\rangle$ obtained via the Dirac quantization with the 
WOC, the latter has the additional term $\frac{1}{8{\cal I}}$ in  (\ref{1stham}). This additional 
contribution originates from the first class constraints in (\ref{tilde12}). The nucleon mass $M_{N}$ ($I=1/2$) 
and delta baryon mass $M_{\Delta}$ ($I=3/2$) are then given as follows~\cite{hong21}
\begin{equation}
M_{N}=ef_{\pi}\left(\frac{6\pi^{2}}{e^{2}}+\frac{e^{2}}{6\pi^{2}}\right),~~~
M_{\Delta}=ef_{\pi}\left(\frac{6\pi^{2}}{e^{2}}+\frac{2e^{2}}{3\pi^{2}}\right).
\label{mnmdelta}
\end{equation}

Next we formulate the magnetic moments of the form~\cite{hong21}
\begin{eqnarray}
\mu_{p}&=&\frac{2M_{N}}{ef_{\pi}}\left(\frac{e^{2}}{48\pi^{2}}+\frac{\pi^{2}}{2e^{2}}\right),~~~~~~~
\mu_{n}=\frac{2M_{N}}{ef_{\pi}}\left(\frac{e^{2}}{48\pi^{2}}-\frac{\pi^{2}}{2e^{2}}\right),\nonumber\\
\mu_{\Delta^{++}}&=&\frac{2M_{N}}{ef_{\pi}}\left(\frac{e^{2}}{16\pi^{2}}+\frac{9\pi^{2}}{10e^{2}}\right),~~~
\mu_{N\Delta}=\frac{2M_{N}}{ef_{\pi}}\cdot\frac{\sqrt{2}\pi^{2}}{2e^{2}},
\label{muspn}
\end{eqnarray}
where $M_{N}$ is now given by the nucleon mass with the WOC in (\ref{mnmdelta}) given in the first class formalism. 
Next we similarly obtain the axial coupling constant~\cite{hong21} 
\begin{equation}g_{A}=\frac{4\pi}{e^{2}}\int_{0}^{\pi}d\mu\sin^{2}\mu\left(1+\cos\mu\right)=\frac{2\pi^{2}}{e^{2}}.
\label{ga1st}
\end{equation}
Now we consider the charge radii. The electric and magnetic isovector mean square charge radii are given in the 
HSM, respectively~\cite{hong98plb,hong21}
\begin{equation}
\langle r^{2}\rangle_{E,I=1}=\langle r^{2}\rangle_{M,I=1}=\frac{2}{3e^{2}{\cal I}}\int_{S^{3}}dV_{B}\sin^{2}\mu\sin^{2}f
\left(1+\left(\frac{d f}{d\mu}\right)^{2}+\frac{\sin^{2}f}{\sin^{2}\mu}\right)=\frac{5}{6e^{2}f_{\pi}^{2}},
\label{misochradius}
\end{equation}
where the subscripts $E$ and $M$ denote electric and magnetic charge radii, respectively, and 
$dV_{B}=\lambda_{B}^{3}\sin^{2}\mu\sin\theta~d\mu~d\theta~d\phi$ on the hypersphere $S^{3}$. Note that 
$dV_{B}$ is given by product of three arc lengths: $\lambda_{B}d\mu$, $\lambda_{B}\sin\mu d\theta$ and 
$\lambda_{B}\sin\mu\sin\theta d\phi$ and $\lambda_{B}$ is radius of hypersphere soliton. Moreover, we find the charge radii in 
terms of $ef_{\pi}$~\cite{hong98plb,hong21}
\begin{eqnarray}
\langle r^{2}\rangle^{1/2}_{M,I=0}&=&\langle r^{2}\rangle^{1/2}_{M,I=1}
=\langle r^{2}\rangle^{1/2}_{M,p}=\langle r^{2}\rangle^{1/2}_{M,n}
=\langle r^{2}\rangle^{1/2}_{E,I=1}=\sqrt{\frac{5}{6}}\frac{1}{ef_{\pi}},\nonumber\\
\langle r^{2}\rangle^{1/2}_{E,I=0}&=&\frac{\sqrt{3}}{2}\frac{1}{ef_{\pi}},~~~
\langle r^{2}\rangle_{p}=\frac{19}{24}\frac{1}{(ef_{\pi})^{2}},~~~
\langle r^{2}\rangle_{n}=-\frac{1}{24}\frac{1}{(ef_{\pi})^{2}}.
\label{chargeradiiall}
\end{eqnarray}
Shuffling the above baryon and transition magnetic moments, we obtain the model independent 
sum rules in the HSM~\cite{hong98plb}
\begin{eqnarray} 
\mu_{\Delta^{++}}&=&\frac{3}{5}(4\mu_{p}+\mu_{n})\nonumber\\
\mu_{N\Delta}&=&\frac{\sqrt{2}}{2}(\mu_{p}-\mu_{n})\nonumber\\
\mu_{p}+\mu_{n}&=&\frac{2}{9}M_{N}(M_{\Delta}-M_{N})\langle r^{2}\rangle_{E,I=0} 
\nonumber\\
\mu_{p}-\mu_{n}&=&\frac{M_{N}}{M_{\Delta}-M_{N}}.
\end{eqnarray}

Next, we choose $\langle r^{2}\rangle_{M,p}^{1/2}=0.80$ fm as an input parameter. We then have 
\begin{eqnarray}
ef_{\pi}=225.23~{\rm MeV}=(0.876~{\rm fm})^{-1},
\label{efpivalue}
\end{eqnarray}
and exploiting this fixed value of $ef_{\pi}$ and the phenomenological formulas 
in (\ref{mnmdelta}) -- (\ref{chargeradiiall}) we can proceed to calculate the physical quantities as shown 
in Table I.

%------------------------------------------------------------------------
%\begin{table}
%\begin{table}[h]
\begin{table}[t]
\caption{In Predictions I and II (Hong)~\cite{hong21}, we use the hypersphere soliton model. 
In Prediction III (ANW), we exploit the standard Skyrmion model. The input parameters are indicated by $*$.}
\begin{center}
\begin{tabular}{crrrr}
\hline
Quantity   &~~~~Prediction I (Hong) &~~~~Prediction II (Hong) &~~~~Prediction III (ANW) &~~~~~~Experiment\\
\hline
$\langle r^{2}\rangle^{1/2}_{M,I=0}$ &0.80 {\rm fm} &0.63 {\rm fm} &0.92 {\rm fm} &0.81 {\rm fm}\\ 
$\langle r^{2}\rangle^{1/2}_{M,I=1}$ &0.80 {\rm fm} &0.63 {\rm fm} &$\infty$ &0.80 {\rm fm}\\ 
$\langle r^{2}\rangle^{1/2}_{M,p}$  &0.80 {\rm fm$^{*}$} &0.63 {\rm fm} &$\infty$ &0.80 {\rm fm}\\
$\langle r^{2}\rangle^{1/2}_{M,n}$  &0.80 {\rm fm}  &0.63 {\rm fm} &$\infty$ &0.79 {\rm fm}\\
$\langle r^{2}\rangle^{1/2}_{E,I=0}$  &0.76 {\rm fm} &0.60 {\rm fm} &0.59 {\rm fm} &0.72 {\rm fm}\\ 
$\langle r^{2}\rangle^{1/2}_{E,I=1}$  &0.80 {\rm fm} &0.63 {\rm fm} &$\infty$  &0.88 {\rm fm}\\
$\langle r^{2}\rangle_{p}$  &(0.780 {\rm fm})$^{2}$ &(0.61 {\rm fm})$^{2}$ &$\infty$  &(0.805 {\rm fm})$^{2}$\\
$\langle r^{2}\rangle_{n}$  &$-(0.179~{\rm fm})^{2}$ &$-(0.14~{\rm fm})^{2}$ &$-\infty$ &$-(0.341~{\rm fm})^{2}$\\
$\mu_{p}$  &2.98 &1.88 &1.87  &2.79\\
$\mu_{n}$  &$-2.45$ &$-1.32$ &$-1.31$ &$-1.91$\\
$\mu_{\Delta^{++}}$  &5.69  &3.72  &3.72 &$4.7-6.7$\\
$\mu_{N\Delta}$  &3.84 &2.27 &2.27 &3.29\\
$M_{N}$ &939 {\rm MeV$^{*}$} &939 {\rm MeV$^{*}$} &939 {\rm MeV$^{*}$} &939 {\rm MeV}\\
$M_{\Delta}$ &1112 {\rm MeV} &1232 {\rm MeV$^{*}$} &1232 {\rm MeV$^{*}$} &1232 {\rm MeV}\\
$g_{A}$ &1.30 &0.98 &0.61 &$1.23$\\
\hline
\end{tabular}
\end{center}
\label{tablestatic}
\end{table}
%---------------------------------------------------------------------------

Now we discuss the predictions in the soliton models. In Table I, Prediction I and II are given by Hong~\cite{hong21} using the HSM, 
while Prediction III is given by ANW~\cite{anw83} exploiting the standard Skyrmion model defined in $R^{3}$ manifold. Here the input 
parameters are indicated by $*$. In Prediction I, the two experimental values for $\langle r^{2}\rangle^{1/2}_{M,p}$ and $M_{N}$ 
are used as input parameters. In Predictions II and III, 
we have exploited the same input parameters associated with $M_{N}$ and $M_{\Delta}$ to compare their predictions effectively. 
Note that in Prediction II we have finite charge 
radii, while in Prediction III we have infinite charge radii. 

Next we discuss the evaluations of Prediction I. 
First, the six predicted values for $\mu_{\Delta^{++}}$, $\langle r^{2}\rangle^{1/2}_{M,I=0}$, 
$\langle r^{2}\rangle^{1/2}_{M,I=1}$, $\langle r^{2}\rangle^{1/2}_{M,n}$ 
(in addition to the input parameters $M_{N}$ and $\langle r^{2}\rangle^{1/2}_{M,p}$) 
are within about 1 \% of the corresponding experimental data. Second, the three predictions for 
$g_{A}$, $\langle r^{2}\rangle^{1/2}_{E,I=0}$ and $\langle r^{2}\rangle_{p}$ are within about 6 \% of the experimental values. 
Third, the three predictions for $M_{\Delta}$, $\mu_{p}$ and $\langle r^{2}\rangle^{1/2}_{E,I=1}$ are within about 10 \% 
of the experimental values.

Now we comment on the hypersurface $A_{3}$ of the hypersphere $S^{3}$ of radius parameter 
$\lambda_{B}$, and the charge radius $\langle r^{2}\rangle_{E,I=1}^{1/2}$ in (\ref{misochradius}). Exploiting the hyperspherical three metric in (\ref{ds2}), 
we find that $A_{3}$ can be analyzed in terms of three arc length elements $\lambda_{B}d\mu$, $\lambda_{B}\sin\mu d\theta$ and $\lambda_{B}\sin\mu\sin\theta d\phi$, 
from which we find the three dimensional hypersurface manifold with $A_{3}=2\pi^{2}\lambda_{B}^{3}$. Note that $\lambda_{B}$ is the radial distance from the center of $S^{3}$ to the hypersphere manifold $S^{3}$ in $R^{4}$. In fact, inserting the value 
$ef_{\pi}=(0.876~{\rm fm})^{-1}$ in  (\ref{efpivalue}) into the condition $L_{B}=1$ in  (\ref{lbequiv}), in the HSM  we obtain the fixed radius parameter given by $\lambda_{B}=\frac{1}{ef_{\pi}}=0.876~{\rm fm}$. On the other hand, the charge radius $\langle r^{2}\rangle_{E,I=1}^{1/2}$ is the physical quantity expressed in (\ref{misochradius}). Integrating over a relevant density on $S^{3}$ corresponding to the integrand in (\ref{misochradius}), we evaluate $\langle r^{2}\rangle_{E,I=1}$ which is now independent of $\mu$, to yield a specific value of the electric isovector root mean square charge radius. The calculated charge radius then can be defined as the fixed radial distance to the point on a hypersurface manifold which does not need to be located only on the compact manifold $S^{3}$ of radius parameter $\lambda_{B}$. This hypersurface manifold is now a submanifold in $R^{4}$ which is located at $r=0.80$ fm far from the center of $S^{3}$.
Note that $\langle r^{2}\rangle_{E,I=1}^{1/2}$ denotes the radial distance which is a geometrical invariant giving the same value both in $R^{3}$ (for instance in volume $R^{3}$ which contains the center of $S^{3}$ and is described in terms of $(r,\theta,\phi)$ at $\mu=\frac{\pi}{2}$) and in $R^{4}$. Next, the physical quantity $\langle r^{2}\rangle_{E,I=1}^{1/2}$ calculated in $R^{3}$ (and in $R^{4}$) then can be compared with the corresponding experimental value, similar to the other physical quantities such as $M_{N}$ and $M_{\Delta}$.

Note that, as a toy model of soliton embedded in $R^{3}$, we consider a uniformly charged manifold 
$S^{2}$ described in terms of $(\theta, \phi)$ and a fixed radius parameter $\lambda_{B}$ 
where we have $A_{2}=4\pi\lambda_{B}^{2}$. By integrating over a surface charge density residing on 
$S^{2}$, one can calculate the physical quantity such as the electric potential, at an arbitrary observation point which does not 
need to be located only on the compact manifold $S^{2}$ of radius parameter $\lambda_{B}$. Next, since the $S^{2}$ soliton of fixed radius parameter $\lambda_{B}$ is embedded in $R^{3}$, we manifestly define an arbitrary radial distance from the center of the 
compact manifold to an observation point which is located in $R^{3}=S^{2}\times R$. Here $S^{2}$ denotes foliation leaves~\cite{foli} of spherical shell of radius parameter $\lambda$ ($0\le\lambda<\infty$) and $R$ is a manifold associated with radial distance.  Note that the radial distance itself is a fixed geometrical invariant producing the same value both in $R^{2}$ (for instance on equatorial 
plane $R^{2}$ which contains the center of $S^{2}$ and is delineated by $(r, \phi)$ at 
$\theta=\frac{\pi}{2}$) and in $R^{3}$. The same mathematical logic can be applied to $S^{3}$ soliton of fixed radius parameter 
$\lambda_{B}$ embedded in $R^{4}=S^{3}\times R$ where $S^{3}$ stands for foliation leaves of hyperspherical shell of radius parameter $\lambda$ ($0\le\lambda<\infty$) and $R$ is a manifold related with radial distance.

Finally we have some comments on the Betti numbers associated with the manifold $S^{3}$ in the HSM. First of all, the $p$-th Betti number $b_{p}(M)$ is defined as the maximal number of $p$-cycles on $M$:
\begin{equation}
b_{p}(M)={\rm dim}~H_{p}(M),
\end{equation}
where $H_{p}(M)$ is the homology group of the manifold $M$~\cite{derham,derham2,derham3}. For the case of $S^{3}$, we obtain
\begin{eqnarray}
H_{0}(S^{3})&=&H_{3}(S^{3})={\mathbf Z},\nonumber\\
H_{p}(S^{3})&=&0,~{\rm otherwise}.
\end{eqnarray}
The non-vanishing Betti numbers related with $S^{3}$ are thus given by $b_{0}(S^{3})=b_{3}(S^{3})=1$.

%%%%%%%%%%%%%%%%%%%%%%%%%%%%%%%%%%%%%%%%%%%%%%%%%%%%%%%%%%%%%%%%%%%%%%%%
\section{Intrinsic Frequencies of Baryons}
\setcounter{equation}{34}
\renewcommand{\theequation}{\arabic{equation}}
%\renewcommand{\theequation}{\arabic{section}.\arabic{equation}}
%%%%%%%%%%%%%%%%%%%%%%%%%%%%%%%%%%%%%%%%%%%%%%%%%%%%%%%%%%%%%%%%%%%%%%%%

Now we investigate the intrinsic frequencies of baryons in the first class Hamiltonian formalism in the 
HSM. To do this, in the HSM we construct 
the equivalent first class Hamiltonian $\tilde{H}^{\prime}$ as follows
\begin{equation}
\tilde{H}^{\prime}=\tilde{H}+\frac{1}{4{\cal I}}\pi_{\theta}\tilde{\Omega}_{2},
\label{equivhtilde}
\end{equation}
to yield the corresponding Gauss law constraint algebra
\begin{equation}\{\tilde{\Omega}_{1},\tilde{H}^{\prime}\}=\frac{1}{2{\cal I}}\tilde{\Omega}_{2},~~~
\{\tilde{\Omega}_{2},\tilde{H}^{\prime}\}=0.
\label{gausslaw}
\end{equation}
Note that $\{\tilde{\Omega}_{1},\tilde{H}\}=0$ and $\{\tilde{\Omega}_{2},\tilde{H}\}=0$. We then find the 
Hamiltonian spectrum for $\tilde{H}^{\prime}$
\begin{equation}\langle \tilde{H}^{\prime}\rangle=E_{B}+\frac{1}{2{\cal I}}\left[I(I+1)+\frac{1}{4}\right],
\label{spectrumprime}
\end{equation}
which is equal to that for $\tilde{H}$ in (\ref{1stham}), as expected. Next we consider the 
equation of motion in Poisson bracket form
\begin{equation}\dot{\tilde{W}}=\{\tilde{W},\tilde{H}^{\prime}\}\nonumber,~~~{\rm for~the~first~class~variable~\tilde{W}},
\label{eomw}
\end{equation}
where the over-dot denotes time derivative. Making use of the equation of motion in (\ref{eomw}), we obtain these two equations
\begin{equation}\dot{\tilde{a}}^{\mu}=\{\tilde{a}^{\mu},\tilde{H}^{\prime}\}=\frac{1}{4{\cal I}}\tilde{\pi}^{\mu},~~~
\dot{\tilde{\pi}}^{\mu}=\{\tilde{\pi}^{\mu},\tilde{H}^{\prime}\}
=-\frac{1}{4{\cal I}}\tilde{\pi}^{\nu}\tilde{\pi}^{\nu}\tilde{a}^{\mu},
\label{twoeom}
\end{equation}
where the first class fields $\tilde{a}^{\mu}$ and $\tilde{\pi}^{\mu}$ are given as follows
\begin{equation}\tilde{a}^{\mu}=a^{\mu}\left(\frac{a^{\nu}a^{\nu}+2\theta}{a^{\nu}a^{\nu}}\right)^{1/2},~~~
\tilde{\pi}^{\mu}=(\pi^{\mu}-a^{\mu}\pi_{\theta})\left(\frac{a^{\nu}a^{\nu}}{a^{\nu}a^{\nu}+2\theta}\right)^{1/2}.
\end{equation}
%--------------to add
In order to formulate the equations in (\ref{twoeom}), we have used the following identities among the physical fields
\begin{eqnarray}
\{\tilde{a}^{\mu},\tilde{\pi}^{\nu}\}&=&\delta^{\mu\nu}-\tilde{a}^{\mu}\tilde{a}^{\nu},~~~~~~~~~~~
\{\tilde{\pi}^{\mu},\tilde{\pi}^{\nu}\}=\tilde{\pi}^{\mu}\tilde{a}^{\nu}-\tilde{a}^{\mu}\tilde{\pi}^{\nu},\nonumber\\
\{\tilde{a}^{\mu},\tilde{H}\}&=&\frac{1}{4{\cal I}}(\tilde{\pi}^{\mu}
-\tilde{a}^{\mu}\tilde{a}^{\nu}\tilde{\pi}^{\nu}),~~~
\{\tilde{\pi}^{\mu},\tilde{H}\}=\frac{1}{4{\cal I}}
(\tilde{\pi}^{\mu}\tilde{a}^{\nu}\tilde{\pi}^{\nu}-\tilde{a}^{\mu}\tilde{\pi}^{\nu}\tilde{\pi}^{\nu}),\nonumber\\
\{\tilde{a}^{\mu},\pi_{\theta}\}&=&\tilde{a}^{\mu},~~~~~~~~~~~~~~~~~~~~~~\{\tilde{\pi}^{\mu},\pi_{\theta}\}=-\tilde{\pi}^{\mu}.
\label{id4} 
\end{eqnarray}
Applying the equation of motion algorithm in (\ref{eomw}) to $\dot{\tilde{a}}^{\mu}$, we find 
\begin{equation}\ddot{\tilde{a}}^{\mu}=\{\dot{\tilde{a}}^{\mu},\tilde{H}^{\prime}\}=\frac{1}{4{\cal I}}\dot{\tilde{\pi}}^{\mu}
=-\frac{1}{4{\cal I}^{2}}\left[I(I+1)+\frac{1}{4}\right]\tilde{a}^{\mu},
\label{ddot1}
\end{equation}
to yield the equation of motion for a simple harmonic oscillator
\begin{equation}
\ddot{\tilde{a}}^{\mu}=-\omega_{I}^{2}\tilde{a}^{\mu},
\label{ddot2}
\end{equation}
where $\omega_{I}$ is the intrinsic frequency of pulsating baryon with isospin quantum number $I$ given by
\begin{equation}
\omega_{I}=\frac{1}{2{\cal I}}\left[I(I+1)+\frac{1}{4}\right]^{1/2}.
\label{omegai}
\end{equation}
Making use of the formula for $\omega_{I}$ in (\ref{omegai}) for the nucleon $N$ $(I=\frac{1}{2})$ and the delta baryon 
$\Delta$ $(I=\frac{3}{2})$, we obtain predictions of intrinsic frequencies $\omega_{N}$ and $\omega_{\Delta}$ 
of the baryons given in Table II. Note that we find the identity $\omega_{\Delta}=2\omega_{N}$.

%------------------------------------------------------------------------
%\begin{table}[p]
\begin{table}[t]
\caption{The intrinsic frequencies of particles~\cite{hong21,hong22}.}
\begin{center}
\begin{tabular}{lll}
\hline
Particle type  &Notation &Intrinsic frequency\\
\hline
Nucleon &$\omega_{N}$ &$0.87\times 10^{23}~{\rm sec}^{-1}$\\
Delta baryon &$\omega_{\Delta}$ &$1.74\times 10^{23}~{\rm sec}^{-1}$\\
Photon &$\omega_{\gamma}$ &$9.00\times 10^{23}~{\rm sec}^{-1}$\\
\hline
\end{tabular}
\end{center}
\label{tableintrinsicfre}
\end{table}
%---------------------------------------------------------------------------

%--------------to add
Finally it seems appropriate to comment on the gauge fixing problem within the first class 
constraints of the Dirac Hamiltonian formalism. In order to investigate the gauge fixing of the first class 
Hamiltonian $\tilde{H}^{\prime}$ in (\ref{equivhtilde}), we introduce 
two canonical sets of ghost and anti-ghost fields together with auxiliary fields
$({\cal C}^{i},\bar{{\cal P}}_{i})$, $({\cal P}^{i},\bar{{\cal C}}_{i})$,
$({\cal N}^{i},{\cal B}_{i})$, $(i=1,2)$ which satisfy the super-Poisson algebra, 
$\{{\cal C}^{i},\bar{{\cal P}}_{j}\}=\{{\cal P}^{i}, \bar{{\cal C}}_{j}\}
=\{{\cal N}^{i},{\cal B}_{j}\}=\delta_{j}^{i}$. Here the super-Poisson bracket is defined as 
$\{A,B\}=\frac{\delta A}{\delta q}|_{r}\frac{\delta B}{\delta p}|_{l}
-(-1)^{\eta_{A}\eta_{B}}\frac{\delta B}{\delta q}|_{r}\frac{\delta A} 
{\delta p}|_{l}$, where $\eta_{A}$ denotes the number of fermions, called the ghost number, in $A$
and the subscript $r$ and $l$ denote right and left derivatives, respectively. 
The BRST charge for the first class constraint algebra related with 
$\tilde{H}^{\prime}$ is then given by
\begin{equation}
Q={\cal C}^{i}\tilde{\Omega}_{i}+{\cal P}^{i}{\cal B}_{i}.
\label{brstcharge}
\end{equation}
We choose the unitary gauge
\begin{equation}
\chi^{1}=\Omega_{1},~~~\chi^{2}=\Omega_{2}, 
\label{unitarygauge}
\end{equation}
by selecting the fermionic gauge fixing function 
$\Psi$: $\Psi=\bar{{\cal C}}_{i}\chi^{i}+\bar{{\cal P}}_{i}{\cal N}^{i}$. 
Exploiting the BRST charge $Q$ in (\ref{brstcharge}), we find the BRST transformation rule defined as $\delta_{Q}F=\{Q,F\}$ 
for a physical field $F$ 
\begin{equation}
\begin{array}{lllll}
\delta_{Q}a^{\mu}=-{\cal C}^{2}a^{\mu}, &\delta_{Q}\pi^{\mu}=2{\cal C}^{1}a^{\mu}+{\cal C}^{2}(\pi^{\mu}-2a^{\mu}\pi_{\theta}),
&\delta_{Q}\theta={\cal C}^{2}a^{\mu}a^{\mu}, &\delta_{Q}\pi_{\theta}=2{\cal C}^{1}, &\delta_{Q}{\cal C}^{i}=0,\\ 
\delta_{Q}\bar{{\cal P}}_{i}=\tilde{\Omega}_{i}, &\delta_{Q}{\cal P}^{i}=0, &\delta_{Q}\bar{{\cal C}}_{i}={\cal B}_{i},
&\delta_{Q}{\cal N}^{i}=-{\cal P}^{i}, &\delta_{Q}{\cal B}_{i}=0.
\end{array}
\label{brsttrfm}
\end{equation}
Note that $\tilde{H}^{\prime}$ is not BRST invariant, which implies that $\delta_{Q}\tilde{H}^{\prime}\neq 0$. 
Next, we obtain the gauge fixed Hamiltonian
\begin{equation}
\tilde{H}^{\prime\prime}=\tilde{H}^{\prime}-\frac{1}{2{\cal I}}{\cal C}^{1}\bar{{\cal P}}_{2},
\label{gaugefixingham}
\end{equation}
which is now invariant under the BRST transformation rule in (\ref{brsttrfm}), namely $\delta_{Q}\tilde{H}^{\prime\prime}=0$.
Note that the BRST charge $Q$ in (\ref{brstcharge}) is nilpotent so that we can have $\delta_{Q^{2}}F=\{Q,\{Q,F\}\}=0$ 
for a physical field $F$. Note also that $H_{eff}\equiv \tilde{H}^{\prime\prime}-\{Q,\Psi\}$ is the BRST invariant Hamiltonian including 
the fermionic gauge fixing function $\Psi$.
%-------------- 

%%%%%%%%%%%%%%%%%%%%%%%%%%%%%%%%%%%%%%%%%%%%%%%%%%%%%%%%%%%%%%%%%%%%%%%%
\section{SPM Predictions}
\setcounter{equation}{48}
\renewcommand{\theequation}{\arabic{equation}}
%\renewcommand{\theequation}{\arabic{section}.\arabic{equation}}
%%%%%%%%%%%%%%%%%%%%%%%%%%%%%%%%%%%%%%%%%%%%%%%%%%%%%%%%%%%%%%%%%%%%%%%%

In this section we will predict the physical quantities such as the photon intrinsic frequency and photon size in the 
SPM~\cite{hong22}. To do this, we will exploit the Nambu-Goto string action~\cite{nambu70,goto71} and its extended rotating bosonic string 
model in $D=3+1$ dimension spacetime~\cite{sato}. 
%--------------to add
Note that in the $D=26$ dimension open string theory which will be briefly discussed below, 
it is well known that there exists the vector boson with 24 independent polarizations~\cite{green87,gova}, 
corresponding to the photon in the stringy photon model defined in the $D=3+1$ dimension spacetime~\cite{hong22} considered in this paper.
%--------------

Before we construct the SPM, we pedagogically summarize a mathematical formalism for the 
Nambu-Goto open string which is related with a photon. 
In order to define the action on curved manifold, we introduce $(M,g_{ab})$ which is a 
$D$ dimensional spacetime manifold $M$ associated with the metric
$g_{ab}$. Given $g_{ab}$, we can have a unique covariant
derivative $\nabla_{a}$ satisfying~\cite{wald84}
\begin{eqnarray}
\nabla_{a}g_{bc}&=&0,\nonumber\\
\nabla_{a}\omega^{b}&=&\partial_{a}\omega^{b}+\Gamma^{b}_{~ac}~\omega^{c},\nonumber\\
(\nabla_{a}\nabla_{b}-\nabla_{b}\nabla_{a})\omega_{c}&=&R_{abc}^{~~~d}~\omega_{d}.\label{rtensor}
\end{eqnarray}

We parameterize an open string by two world sheet coordinates
$\tau$ and $\sigma$, and then we have the corresponding vector
fields $\xi^{a}=(\partial/\partial\tau)^{a}$ and
$\zeta^{a}=(\partial/\partial\sigma)^{a}$.  The Nambu-Goto string action is
now given by~\cite{nambu70,goto71}
\begin{equation} 
S=-\kappa\int\int~d\tau
d\sigma f(\tau,\sigma),\label{nambugoto}
\end{equation} 
where the coordinates $\tau$ and
$\sigma$ have ranges $\tau_{1}\leq \tau\leq \tau_{2}$ and $0\leq \sigma\leq
\pi$ respectively and
\begin{equation}
f(\tau,\sigma)=[(\xi\cdot\zeta)^{2}-(\xi\cdot\xi)(\zeta\cdot\zeta)]^{1/2}.
\label{fts0}
\end{equation}
Here the string tension $\kappa$ is defined by $\kappa=\frac{1}{2\pi \alpha^{\prime}}$, with $\alpha^{\prime}$ being the universal 
slope of the linear Regge trajectories~\cite{scherk75}. 

We perform an infinitesimal variation of the world sheets
$\gamma_{\alpha}(\tau,\sigma)$ traced by the open string during
its evolution in order to find the string geodesic equation from 
least action principle.  Here we impose the restriction that
the length of the string is $\tau$ independent. We introduce the deviation vector 
$\eta^{a}=(\partial/\partial \alpha)^{a}$ which represents the displacement to an infinitesimally
nearby world sheet, and we consider $\Sigma$ which denotes the three dimensional
submanifold spanned by the world sheets $\gamma_{\alpha}(\tau,\sigma)$.
We then may choose $\tau$, $\sigma$ and $\alpha$ as coordinates of
$\Sigma$ to yield the commutator relations 
\begin{eqnarray}
\pounds_{\xi}\eta^{a}&=&\xi^{b}\nabla_{b}\eta^{a}-\eta^{b}\nabla_{b}\xi^{a}=0,\nonumber\\
\pounds_{\zeta}\eta^{a}&=&\zeta^{b}\nabla_{b}\eta^{a}-\eta^{b}\nabla_{b}\zeta^{a}=0,\nonumber\\
\pounds_{\xi}\zeta^{a}&=&\xi^{b}\nabla_{b}\zeta^{a}-\zeta^{b}\nabla_{b}\xi^{a}=0.\label{poundxizeta}
\end{eqnarray} 

Now we find the first variation as follows
\begin{equation}
\frac{dS}{d\alpha} =\int\int d\tau
d\sigma~\eta_{b}(\xi^{a}\nabla_{a}p^{b}+\zeta^{a}\nabla_{a}\pi^{b})
-\int d\sigma~p^{b}\eta_{b}|_{\tau=\tau_{1}}^{\tau=\tau_{2}}-\int
d\tau~\pi^{b}\eta_{b}|_{\sigma=0}^{\sigma=\pi},\label{dsdalpha2}
\end{equation} 
where the world sheet currents associated with $\tau$ and
$\sigma$ directions are respectively given by~\cite{scherk75}, 
\begin{eqnarray}
p^{a}&=&\frac{\kappa}{f}[(\xi\cdot\zeta)\zeta^{a}-(\zeta\cdot\zeta)\xi^{a}],\nonumber\\
\pi^{a}&=&\frac{\kappa}{f}[(\xi\cdot\zeta)\xi^{a}-(\xi\cdot\xi)\zeta^{a}].
\label{pps2}\end{eqnarray}
Using the endpoint conditions 
\begin{equation}
\eta^{a}(\tau=\tau_{1};\sigma)=\eta^{a}(\tau=\tau_{2};\sigma)=0,
\end{equation} 
and 
\begin{equation}
\pi^{a}(\tau;\sigma=0)=\pi^{a}(\tau;\sigma=\pi)=0,
\label{end}
\end{equation} 
we have string geodesic equation
\begin{equation}
\xi^{a}\nabla_{a}p^{b}+\zeta^{a}\nabla_{a}\pi^{b}=0,\label{geodesicng0}
\end{equation}
and constraint identities~\cite{scherk75} 
\begin{eqnarray}
p\cdot\zeta&=&0,~~~p\cdot p+\kappa^{2}\zeta\cdot\zeta=0,\nonumber\\
\pi\cdot\xi&=&0,~~~\pi\cdot \pi+\kappa^{2}\xi\cdot\xi=0.
\label{consts2} 
\end{eqnarray}
%--------------to add
For more details of the string theory and deviation vector on the curved manifold, 
see the references~\cite{hongcho08,hongcho11,wald84}.
%--------------

Next we consider the open rotating string in the (3+1) dimensional flat 
spacetime and delineate the string in terms of the coordinates
\begin{equation}
x_{\mu}=(x_{0},x_{i})=(\tau,x_{i}(\tau;\sigma)),~~~(i=1,2,3).
\label{xmu123}
\end{equation}
The Nambu-Goto string action in (\ref{nambugoto}) is then described in terms of $f(\tau,\sigma)$ given by
\begin{equation}
f(\tau,\sigma)=[(\dot{x}_{\mu}{x}^{\prime}_{\mu})^{2}-(\dot{x}_{\mu}\dot{x}_{\mu})({x}^{\prime}_{\mu}{x}^{\prime}_{\mu})]^{1/2},
\label{fts0}
\end{equation}
where the overdot and prime denote derivatives with respect to $\tau$ and $\sigma$, respectively. In this paper, we 
use the metric signature $(+, -, -, -)$.

Inserting (\ref{xmu123}) into (\ref{fts0}), we find 
\begin{equation}
f(\tau,\sigma)=[(\dot{x}_{i}{x}^{\prime}_{i})^{2}+(1-\dot{x}_{i}\dot{x}_{i})x^{\prime}_{j}x^{\prime}_{j}]^{1/2}.
\label{fts}
\end{equation}
Moreover we proceed to construct the world sheet currents
\begin{eqnarray}
p_{0}&=&\frac{\kappa}{f}x^{\prime}_{i}x^{\prime}_{i},~~~~~p_{i}=-\frac{\kappa}{f}[(\dot{x}_{j}x^{\prime}_{j})x^{\prime}_{i}-(x^{\prime}_{j}
x^{\prime}_{j})\dot{x}_{i}],\nonumber\\
\pi_{0}&=&-\frac{\kappa}{f}\dot{x}_{i}x^{\prime}_{i},~~~\pi_{i}=-\frac{\kappa}{f}[(\dot{x}_{j}x^{\prime}_{j})\dot{x}_{i}+(1-\dot{x}_{j}\dot{x}_{j})x^{\prime}_{i}].
\label{pppp}
\end{eqnarray}
Now, exploiting (\ref{geodesicng0}) we obtain the string equation of motion
\begin{equation}
\frac{\partial p_{\mu}}{\partial \tau}+\frac{\partial \pi_{\mu}}{\partial \sigma}=0,
\label{geodesicng}
\end{equation}
and the string boundary condition
\begin{equation}
\pi_{\mu}(\tau;\sigma=0)=\pi_{\mu}(\tau;\sigma=\pi)=0.
\label{end}
\end{equation}
Inserting $p_{\mu}$ and $\pi_{\mu}$ in (\ref{pppp}) into the string equation of motion
in (\ref{geodesicng}), we find
\begin{eqnarray}
\frac{\partial}{\partial\tau}\left(\frac{x^{\prime}_{i}x^{\prime}_{i}}{f}\right)
-\frac{\partial}{\partial\sigma}\left(\frac{\dot{x}_{i}x^{\prime}_{i}}{f}\right)&=&0,\nonumber\\
\frac{\partial}{\partial\tau}\left[\frac{(\dot{x}_{j}x^{\prime}_{j})x^{\prime}_{i}-(x^{\prime}_{j}x^{\prime}_{j})\dot{x}_{i}}{f}\right]
+\frac{\partial}{\partial\sigma}\left[\frac{(\dot{x}_{j}x^{\prime}_{j})\dot{x}_{i}+(1-\dot{x}_{j}\dot{x}_{j})x^{\prime}_{i}}{f}\right]&=&0.
\label{eom2ns}
\end{eqnarray}
Exploiting the boundary conditions in (\ref{end}), we also obtain at $\sigma=0$ and $\sigma=\pi$
\begin{equation}
\dot{x}_{i}x^{\prime}_{i}=0,~~~(1-\dot{x}_{j}\dot{x}_{j})x^{\prime}_{i}=0.
\label{end2}
\end{equation}

Next, in order to describe an open string, which is rotating in $(x_{1},x_{2})$ plane and residing 
on the string center of mass, we take an ansatz~\cite{sato}
\begin{equation}
x_{i}^{rot}=(r(\sigma)\cos\omega\tau,r(\sigma)\sin\omega\tau,0).
\label{xrot}
\end{equation} 
Here we propose that $r(\sigma)$ and $\omega$ represent respectively the diameter and angular velocity 
of the photon with solid spherical shape which is delineated by the open string. Note that $r(\sigma=\pi/2)$ denotes the center 
of the diameter of string. More specifically, $r(\sigma=\pi/2)$ is located in the center of the solid sphere 
which describes the photon. The first boundary condition 
in (\ref{end2}) is trivially satisfied and the second one yields
\begin{equation}
r^{\prime}(\sigma=0,\pi)=0.
\label{bc}
\end{equation}
We then obtain $r(\sigma)$ which fulfills the above condition in (\ref{bc}) 
\begin{equation}
r(\sigma)=\frac{1}{\omega}\cos\sigma.
\label{romega}
\end{equation} 
Note that the photon has a finite size which is filled with mass. 
Using the photon configuration in (\ref{xrot}) and (\ref{romega}) together with (\ref{pppp}), we find 
the rotational energy of the photon
\begin{equation}
E^{rot}=\int_{0}^{\pi}d\sigma~p_{0}^{rot}=\frac{1}{2\alpha^{\prime}\hbar\omega},
\label{hrot}
\end{equation}
where we have included $\hbar$ factor explicitly, and the value of $\alpha^{\prime}$ is given by 
$\alpha^{\prime}=0.95~{\rm GeV}^{-2}$~\cite{scherk75}. 

Next we evaluate the photon intrinsic frequency and size in the SPM. To do this, we 
calculate the vibrational energy of photon by introducing the string 
coordinate configurations
\begin{equation}
x_{i}=x_{i}^{rot}+y_{i},~~~i=1,2,3.
\label{yi}
\end{equation} 
Exploiting the coordinates in (\ref{yi}), we expand the string Lagrangian density
\begin{equation}
{\cal L}={\cal L}_{0}+\frac{1}{2}\frac{\partial^{2}{\cal L}}{\partial\dot{x}_{i}\partial\dot{x}_{j}}|_{0}\dot{y}_{i}\dot{y}_{j}
+\frac{\partial^{2}{\cal L}}{\partial\dot{x}_{i}\partial x^{\prime}_{j}}|_{0}\dot{y}_{i}y^{\prime}_{j}
+\frac{1}{2}\frac{\partial^{2}{\cal L}}{\partial x^{\prime}_{i}\partial x^{\prime}_{j}}|_{0}y^{\prime}_{i}y^{\prime}_{j}+\cdots,
\label{call}
\end{equation}
where the subscript $0$ denotes that the terms in (\ref{call}) are evaluated by using the coordinates 
in (\ref{xrot}). The ellipsis stands for the higher derivative terms. Here the first term is a constant given by 
${\cal L}_{0}={\cal L}(x_{i}^{rot})$. The first derivative terms vanish after exploiting the string equation of motion in 
(\ref{geodesicng}). Next in order to obtain the 
vibration energy of photon, 
we define coordinates $z_{i}$ which co-rotates with the string itself
\begin{eqnarray}
z_{1}&=&y_{1}\cos\omega\tau+y_{2}\sin\omega\tau,\nonumber\\
z_{2}&=&-y_{1}\sin\omega\tau+y_{2}\cos\omega\tau,\nonumber\\
z_{3}&=&y_{3}.
\label{zzz}
\end{eqnarray}  
After some algebra, we obtain the Lagrangian density associated with the coordinates $z_{i}$
\begin{eqnarray}
{\cal L}(z_{i})&=&\frac{\kappa}{2\sin^{2}\sigma}\left[\frac{1}{\omega}(\dot{z}_{2}+\omega z_{1})^{2}
+2\sin\sigma\cos\sigma((\dot{z}_{1}-\omega z_{2})z_{2}^{\prime}\right.\nonumber\\
&&\left.-(\dot{z}_{2}+\omega z_{1})z_{1}^{\prime})
-\omega z^{\prime 2}_{2}\right]+\frac{\kappa}{2\omega}(\dot{z}_{3}^{2}-\omega^{2}z_{3}^{\prime 2}).
\label{lagz}
\end{eqnarray}

The equations of motion for the directions $z_{2}$ and $z_{3}$ are then given by
\begin{eqnarray}
\ddot{z}_{2}+\omega^{2}z_{2}+2\omega^{2}\cot\sigma z_{2}^{\prime}-\omega^{2} z_{2}^{\prime\prime}&=&0,\nonumber\\
\ddot{z}_{3}-\omega^{2}z_{3}^{\prime\prime}&=&0.\label{z3eom}
\end{eqnarray} 
Now the photon is assumed to be in the ground state of the string energy spectrum. From (\ref{z3eom}) we find the 
eigenfunctions for the ground states
\begin{eqnarray}
{z}_{2}&=&c_{2}\sin(\omega\tau+\phi_{2}),\label{z2sol}\\
{z}_{3}&=&c_{3}\cos\sigma\sin(\omega\tau+\phi_{3}).\label{z3sol}
\end{eqnarray}
Here $\phi_{2}$ and $\phi_{3}$ are arbitrary phase constants which are irrelevant to the physics arguments of interest. 

It seems appropriate to address comments on the photon vibration modes. The transverse mode $z_{2}$ in (\ref{z2sol}) 
is independent of the string coordinate $\sigma$, so that the photon can tremble back and forth with 
a constant amplitude, while the longitudinal mode $z_{3}$ in (\ref{z3sol}) possesses 
sinusoidal dependence on $\sigma$. Here note that $z_{3}$ does not move at the center of the string, namely at $\sigma=\pi/2$, 
independent of $\tau$ and the other parts of the string oscillate with the frequency $\omega$. As for the transverse mode $z_{1}$, 
we can find that any value for $z_{1}$ satisfies the Euler-Lagrange equation for $z_{1}$ obtained from 
the Lagrangian density in (\ref{lagz}). Up to now we have considered a single massive photon with the 
solid sphere shape, whose diameter is delineated in terms of the length of the open string. The photon thus has a disk-like cross section 
on which the coordinates $z_{1}$ and $z_{2}$ resides. Note that, similar to the phonon associated with massive particle 
lattice vibrations, the photon is massive so that we can have three polarization directions: two transverse directions 
as in the massless photon case, and an additional longitudinal one. Keeping this argument in mind, we find that there 
exist two transverse modes $z_{1}$ and $z_{2}$ associated with the photon vibrations on $z_{1}$-$z_{2}$ 
cross sectional disk, in addition to one longitudinal mode $z_{3}$. We thus have the transverse mode in $z_{1}$ direction 
to yield the eigenfunction for the ground state, with an arbitrary phase constant $\phi_{1}$ similar to $\phi_{2}$ and $\phi_{3}$ discussed above,
\begin{equation}
{z}_{1}=c_{1}\sin(\omega\tau+\phi_{1}).\label{z1sol}
\end{equation}
Note that, as in the case of massless photon, $z_{1}$ mode oscillates with the same frequency $\omega$ that $z_{2}$ mode does. 
Note also that the above solutions $z_{i}$ satisfy their endpoint conditions at $\sigma=0$ and $\sigma=\pi$
\begin{equation}
z_{i}^{\prime}=0,~~~i=1,2,3.
\end{equation}
The energy eigenvalues in the ground states in (\ref{z2sol})--(\ref{z1sol}) are then given by
\begin{equation}
E^{vib}_{i}=\frac{1}{2}\hbar \omega,~~~i=1,2,3.\label{evib123}
\end{equation}
Exploiting the energies in (\ref{evib123}), we arrive at the vibrational energy of the open string ground state
\begin{equation}
E^{vib}=\sum_{i=1}^{3}E^{vib}_{i}=\frac{3}{2}\hbar\omega.\label{evibtot}
\end{equation}

In the SPM the classical energy of the string is 
given by $E^{rot}$ in (\ref{hrot}), while the corresponding quantum mechanical energy is given by $E^{vib}$ in (\ref{evibtot}). 
Now we define the total energy of the string configuration as a function of 
$\omega$
\begin{equation}
E=E^{rot}+E^{vib}=\frac{1}{2\alpha^{\prime}\hbar\omega}+\frac{3}{2}\hbar\omega.
\label{stringenergyrv}
\end{equation}
Note that we have already removed the translational degree of freedom, by considering the string observer 
residing on the photon center of mass in the SPM. Variating the energy $E$ with respect to $\omega$, we 
find the minimum value condition for $E$ at $\omega=\omega_{\gamma}$
\begin{equation}\frac{dE}{d\omega}(\omega=\omega_{\gamma})=0,
\label{dedomega}
\end{equation}
to yield the intrinsic frequency $\omega_{\gamma}$ for the photon~\cite{hong22}
\begin{equation}\omega_{\gamma}=\frac{1}{\hbar(3\alpha^{\prime})^{1/2}}=9.00\times 10^{23}~{\rm sec}^{-1},
\label{omegagammasec}
\end{equation}
which is graeter than the baryon intrinsic frequencies as shown in Table II. 
%--------------to add
Note that we have related the HSM and SPM, in terms of the intrinsic 
frequencies of the baryons and photon both of which are the extended objects.
%--------------
Next we consider the photon radius as a 
half of the open string length. Exploiting the photon intrinsic frequency $\omega_{\gamma}$ in (\ref{omegagammasec}) 
we obtain the photon radius~\cite{hong22} 
\begin{equation}\langle r^{2}\rangle^{1/2}(\rm photon)=\frac{c}{2\omega_{\gamma}}=0.17~{\rm fm},
\label{photonradius}
\end{equation}
which is 21\% of the proton magnetic charge radius $\langle r^{2}\rangle^{1/2}_{M}(\rm proton)=0.80~{\rm fm}$ as shown in Table I.

%--------------to add
Next, even though up to now we have investigated the stringy photon model in $D=3+1$ dimension spacetime~\cite{hong22}, 
we parenthetically discuss the bosonic string theory in the critical dimension $D=26$.
%--------------
In the light-cone gauge quantization in bosonic string theory, the so-called anomaly associated with 
commutator of Lorentz generators has been canceled in $D=26$ critical dimensions and with the condition 
$a=-\frac{D-2}{2}R~=1$~\cite{green87,gova}, where $R$ is the Ramanujan evaluation~\cite{berndt85} for the infinite sum
\begin{equation}R\equiv 1+2+3+\cdots=-\frac{1}{12}.
\label{ramaequiv}
\end{equation}
Now we investigate the Ramanujan evaluation procedure for $R$. To accomplish this, we manipulate the difference $R-4R$ to yield 
$R-4R=-3R=1+(2-4)+3+(4-8)+5+\cdots=1-2+3-4+5+\cdots=\frac{1}{4}$. 
Here we have used the identity $1+2x+3x^{2}+4x^{3}+\cdots=\frac{1}{(1-x)^{2}}$ 
which yields $1-2+3-4+\cdots=\frac{1}{4}$ at $x=-1$. Following the above relation $R-4R=-3R=\frac{1}{4}$, we finally obtain 
the Ramanujan evaluation $R=-\frac{1}{12}$ in (\ref{ramaequiv}). 
Next exploiting the Ramanujan evaluation in (\ref{ramaequiv}) we can obtain the relation 
\begin{equation}
\frac{1}{2}+\frac{3}{2}+\frac{5}{2}+\cdots=\frac{1}{2}R-R=\frac{1}{24},
\label{ramansuper}
\end{equation}
which has been used in the $D=10$ superstring theory~\cite{gova,hong22}. 
%--------------to add
Note that the stringy photon model has been described in $D=3+1$ 
dimension spacetime, without resorting to (\ref{ramaequiv}) and (\ref{ramansuper}).
%--------------

%%%%%%%%%%%%%%%%%%%%%%%%%%%%%%%%%%%%%%%%%%%%%%%%%%%%%%%%%%%%%%%%%%%%%%%%
\section{Conclusions}
%\setcounter{equation}{0}
%\renewcommand{\theequation}{\arabic{section}.\arabic{equation}}
%%%%%%%%%%%%%%%%%%%%%%%%%%%%%%%%%%%%%%%%%%%%%%%%%%%%%%%%%%%%%%%%%%%%%%%%

In summary, we have formulated the HSM to discuss the Dirac quantization in the first class formalism and the 
predictions of baryon physical quantities in HSM. To be specific, we have evaluated the intrinsic pulsating 
frequencies of the baryons. To accomplish this, we have exploited the first class Hamiltonian possessing the WOC and 
quantized the hypersphere soliton in the HSM. Here we have noticed that the the 
profile function in the soliton energy of the hypersphere soliton satisfies the two first order differential equations 
to attain the BPS topological lower bound in the soliton energy. Next, we have evaluated the baryon physical quantities 
such as baryon masses, magnetic moments, charge radii and axial coupling constant. Shuffling the baryon 
and transition magnetic moments, we have constructed the model independent sum rules. The intrinsic frequency for 
more massive particle also has been shown to be greater than that for the less massive one. 
Explicitly we have evaluated the intrinsic frequencies $\omega_{N}=0.87\times 10^{23}~{\rm sec}^{-1}$ and 
$\omega_{\Delta}=1.74\times 10^{23}~{\rm sec}^{-1}$ of the nucleon and delta baryon, respectively, 
to yield the identity $\omega_{\Delta}=2\omega_{N}$.

Next, making use of the Nambu-Goto string action and its extended rotating bosonic string theory, 
we have formulated the SPM to find the energy of the string configuration, 
which consists of the rotational and vibrational energies of the open string. We have then investigated a 
photon intrinsic frequency by exploiting the string which performs both rotational 
and pulsating motions. Here we have interpreted the string as a diameter of a solid spherical shape photon. 
Explicitly we have found that the intrinsic frequency of the photon is comparable to those of the baryons such as 
nucleon and delta baryon. In the SPM we also have evaluated the photon size given by the string radius which is approximately 21\% of 
the proton magnetic charge radius. It will be interesting to search for a strong experimental evidence for the photon size which 
could be associated with the manifest photon phenomenology such as the photoelectric effect, Compton scattering and 
Raman scattering, for instance. Assuming that the SPM exploited in this paper could 
be a precise description for the photon, the photon intrinsic frequency 
$\omega_{\gamma}=9.00\times 10^{23}~{\rm sec}^{-1}$ and photon size $\langle r^{2}\rangle^{1/2}(\rm photon)=0.17~{\rm fm}$ 
could be fundamental predictions in the extended object phenomenology, similar to $\omega_{N}$, $\omega_{\Delta}$ and the charge radii 
in the hypersphere soliton model.\\ \\ \\

\noindent{\bf Acknowledgments}\\ 

The author would like to thank the anonymous editor and referees for helpful comments.\\ \\

\noindent{\bf Funding}\\ 

S.T. Hong was supported by Basic Science Research Program through 
the National Research Foundation of Korea funded by the Ministry of Education, NRF-2019R1I1A1A01058449.\\ \\

\noindent{\bf Data Availability Statement}\\ 

No data has been used in the work.\\ \\

\noindent{\bf Conflicts of Interest}\\ 

The author declares no conflict of interest.\\ \\ \\

%%%%%%%%%%%%%%%%%%%%%%%%%%%%%%%%%%%%%%%%%%%%%%%%%%%%%%%%%


\begin{thebibliography}{99}
\bibitem{skyrme61} Skyrme, T.H.R. A nonlinear field theory. {\it Proc. R. Soc. Lond. A} {\bf 1961}, 260, 127.
\bibitem{faddeev} Faddeev, L.D. Some comments on the many dimensional solitons. {\it Lett. Math. Phys.} {\bf 1976}, 1, 289.
\bibitem{anw83} Adkins, G.S.; Nappi, C.R.; Witten, E. Static properties of nucleons in the Skyrme model. {\it Nucl. Phys. B} {\bf 1983}, 
228, 552.
\bibitem{manton1} Manton, N.S.; Ruback, P.J. Skyrmions in flat space and curved space. {\it Phys. Lett. B} {\bf 1986}, 181, 137.
\bibitem{liu87} Adkins, G.S. {\it Chiral Solitons}; Liu, K.F., ED.; World Scientific: Singapore, 1987.
\bibitem{hong98plb} Hong, S.T. The static properties of Skyrmions on a hypersphere. {\it Phys. Lett. B} {\bf 1998}, 417, 211 
[Corrigendum in {\it Phys. Lett. B} {\bf 2020}, 803, 135179].
\bibitem{manton2} Manton, N.S.; Sutcliffe, P. {\it Topological Solitons}; Cambridge University Press: Cambridge, UK, 2004.
\bibitem{hong15} Hong, S.T. {\it BRST Symmetry and de Rham Cohomology}; Springer: Heidelberg, Germany, 2015.
\bibitem{hong21} Hong, S.T. Dirac quantization and baryon intrinsic frequencies in hypersphere soliton model. {\it Nucl. Phys. B} 
{\bf 2021}, 973, 115611.
\bibitem{green87} Green, M.B.; Schwarz, J.H.; Witten, E. {\it Superstring Theory}; Cambridge University Press: Cambridge, UK, 1987.
\bibitem{gova} Govaerts, J. The geometrical principles of string theory constructions. In 
{\it Particle Physics Carg\`ese 1989}; M. Levy, M., Basdevant, J.L., Jacob, M., Speiser, D., Weyers, J., Gastmans, R., Eds.; 
Springer: Heidelberg, Germany, 1990.
\bibitem{polchinski98} Polchinski, J. {\it String Theory}; Cambridge University Press: Cambridge, UK, 1998.
\bibitem{hong22} Hong, S.T. Photon intrinsic frequency and size in stringy photon model. {\it Nucl. Phys. B} {\bf 2022}, 976, 115720.
\bibitem{di62} Dirac, P.A.M. An extensible model of the electron. {\it Proc. Roy. Soc. Lond. A} {\bf 1962}, 268, 57.
\bibitem{hongcho08} Cho, Y.S.; Hong, S.T. Singularities in geodesic surface congruence. {\it Phys. Rev. D} {\bf 2008}, 78, 067301.
\bibitem{hongcho11} Cho, Y.S.; Hong, S.T. Dynamics of stringy congruence in early universe. {\it Phys. Rev. D} {\bf 2011}, 83, 104040.
\bibitem{nambu70} Nambu, Y. Lecture at the Copenhagen Symposium. University of Chicago, Chicago, IL, USA, 1970, unpublished.
\bibitem{goto71} Goto, T. Relativistic quantum mechanics of one-dimensional mechanical continuum and subsidiary condition 
of dual resonance model. {\it Prog. Theor. Phys.} {\bf 1971}, 46, 1560.
\bibitem{schwarz74} Scherk, J.; Schwarz, J.H. Dual models and the geometry of space-time. {\it Phys. Lett. B} {\bf 1974}, 52, 347.
\bibitem{manton3} Manton, N.S. Geometry of Skyrmions. {\it Commun. Math. Phys.} {\bf 1987}, 111, 469.
\bibitem{jackson88} Jackson, A.D.; Wirzba, A.; Castillejo, L. Two Skyrmions on a hypersphere. {\it Nucl. Phys. A} {\bf 1988}, 486, 634.
\bibitem{di} Dirac, P.A.M. {\it Lectures on Quantum Mechanics}; Yeshiva University Press: New York, NY, USA, 1964.
\bibitem{lee81} Lee, T.D. {\it Particle Physics and Introduction to Field Theory}; 
Harwood Academic Publishers: London, UK, 1981; p. 477.
\bibitem{neto95} Neto, J.A. Remarks on the collective quantization of the SU(2) Skyrmion model. 
{\it J. Phys. G} {\bf 1995}, 21, 695.
\bibitem{foli} Candel, A.; Conlon, L. {\it Foliations I}; American Mathematical Society: Providence, RI, USA, 2000.
\bibitem{derham} Rotman, J.J. {\it An Introduction to Algebraic Topology}; Springer: Heidelberg, Germany, 1988.
\bibitem{derham2} Weibel, C.A. {\it An Introduction to Homological Algebra}; Cambridge University Press: Cambridge, UK, 1994.
\bibitem{derham3} Frankel, T.  {\it The Geometry of Physics}; Cambridge University Press: Cambridge, UK, 1997.
\bibitem{sato} Kikkawa, K.; Sato, M.; Uehara, K. Non-linear strings. {\it Prog. Theor. Phys.} {\bf 1977}, 57, 2101.
\bibitem{wald84} Wald, R.M. {\it General Relativity}; University of Chicago Press: Chicago, IL, USA, 1984.
\bibitem{scherk75} Scherk, J. An introduction to the theory of dual models and strings. {\it Rev. Mod. Phys.} {\bf 1975}, 47, 123.
\bibitem{berndt85} Berndt, B.C. {\it Ramanujan's Notebooks: Part 1}; Springer: Heidelberg, Germany, 1985; p. 135.
\end{thebibliography}
\end{document}